\documentclass[conference]{IEEEtran}
\ifCLASSINFOpdf
\else
\fi

\usepackage{algorithm}
\usepackage{algpseudocode}
\usepackage{subfigure}
\usepackage{graphicx}

\usepackage{amsmath}
\usepackage{amssymb}

\begin{document}

\title{Optimizing Distributed Training Approaches for Scaling Neural Networks}


\author{
\IEEEauthorblockN{\IEEEauthorrefmark{1}{ Vishnu Vardhan Baligodugula}, 
\IEEEauthorrefmark{1}$^{}$Fathi Amsaad\thanks{* Corresponding author}
}
\IEEEauthorblockA{
\IEEEauthorrefmark{2} Department of Computer Science and Engineering, Wright State University\\
Email: \IEEEauthorrefmark{2}\{baligodugula.2,fathi.amsaad\}@wright.edu\\
}
}


%


\maketitle

\begin{abstract}

This paper presents a comparative analysis of distributed training strategies for large-scale neural networks, focusing on data parallelism, model parallelism, and hybrid approaches. We evaluate these strategies on image classification tasks using the CIFAR-100 dataset, measuring training time, convergence rate, and model accuracy. Our experimental results demonstrate that hybrid parallelism achieves a 3.2× speedup compared to single-device training while maintaining comparable accuracy. We propose an adaptive scheduling algorithm that dynamically switches between parallelism strategies based on network characteristics and available computational resources, resulting in an additional 18\% improvement in training efficiency.

\end{abstract}


%
\IEEEpeerreviewmaketitle

\section{Introduction}

The increasing size and complexity of deep neural networks have made distributed training essential for modern machine learning applications. While models continue to grow in parameter count and computational requirements, training efficiency has become a critical bottleneck in deep learning research and deployment. State-of-the-art models often require days or weeks of training on specialized hardware, making experimentation costly and time-consuming.

Distributed training approaches address this challenge by parallelizing computation across multiple devices. However, the optimal parallelization strategy depends on various factors including model architecture, dataset characteristics, and hardware configuration. Current approaches typically employ fixed parallelism strategies that may not adapt well to different phases of training or to different components of a neural network.

In this paper, we make the following contributions:

\begin{itemize}
    \item We provide a systematic comparison of data parallelism, model parallelism, and hybrid approaches across different model architectures and scales.
    \item We propose a novel adaptive scheduling algorithm that dynamically selects the optimal parallelism strategy for different components of a neural network.
    \item We demonstrate significant improvements in training efficiency without sacrificing model accuracy through extensive experimentation on standard benchmarks.
\end{itemize}

\section{Related Work}

\subsection{Data Parallelism}
Data parallelism replicates the model across multiple devices, with each device processing different batches of training data. Goyal et al. \cite{c1} demonstrated the effectiveness of large-batch training with data parallelism, achieving linear scaling on ImageNet training by adjusting the learning rate. However, data parallelism faces communication bottlenecks due to gradient synchronization, especially as the number of devices increases \cite{c2}.

\subsection{Model Parallelism}
Model parallelism partitions the neural network across devices, allowing training of models that exceed single-device memory constraints. Megatron-LM \cite{c3} implemented tensor parallelism for large language models by partitioning individual layers. Pipeline parallelism \cite{c4} divides models into sequential stages processed by different devices, though this approach introduces pipeline bubbles that reduce hardware utilization.

\subsection{Hybrid Approaches}
Recent work has explored combining different parallelism strategies. Shoeybi et al. \cite{c5} combined data and model parallelism for training language models with billions of parameters. These hybrid approaches have shown promise but typically employ static partitioning schemes that may not adapt to changing computational patterns during training.

\subsection{Adaptive Training Strategies}
Adaptive methods have been explored for various aspects of deep learning, including learning rates \cite{c6} and batch sizes \cite{c7}. However, adaptive parallelism strategies remain relatively unexplored. FlexFlow \cite{c8} proposed an optimization framework for finding efficient parallelization strategies but does not adapt during training.

\section{Methodology}

\subsection{System Architecture}
Our distributed training framework consists of a coordinator node and multiple worker nodes. The coordinator manages the training process, including parallelism strategy selection and synchronization. Each worker executes assigned computations and communicates with other workers as needed. The framework supports three primary parallelism strategies:

\begin{itemize}
    \item \textbf{Data Parallelism (DP):} The complete model is replicated across all devices, with each device processing a different subset of the training data. Gradients are synchronized across devices after each forward-backward pass.
    \item \textbf{Model Parallelism (MP):} The model is partitioned across devices, with each device responsible for a specific portion of the computation graph. This approach is particularly beneficial for models that exceed single-device memory constraints.
    \item \textbf{Hybrid Parallelism (HP):} Combines both data and model parallelism, applying different strategies to different parts of the network based on their computational and memory requirements.
\end{itemize}

\subsection{Adaptive Scheduling Algorithm}
The core contribution of our work is an Adaptive Scheduling Algorithm \ref{algori} (ASA) that dynamically selects the optimal parallelism strategy for each component of the neural network. The algorithm operates as follows:

\begin{enumerate}
    \item The neural network is partitioned into logical components (e.g., individual layers or blocks).
    \item During an initial profiling phase, the execution time and memory requirements of each component are measured.
    \item The communication overhead for different parallelism strategies is estimated based on the size of activations and gradients.
    \item An optimization problem is solved to minimize the overall training time, considering both computation and communication costs.
    \item The selected parallelism strategy is applied for each component.
    \item The profiling and strategy selection are periodically updated during training to adapt to changing patterns.
\end{enumerate}

\begin{algorithm}
\caption{Training with Parallelism Strategy Optimization}
\label{algori}
\begin{algorithmic}[1]
\State \textbf{Input:} Model $M$, Dataset $D$, Available Devices $K$
\State \textbf{Output:} Trained model parameters $\theta$, Training time $T$
\State Initialize model parameters $\theta$ randomly
\State Partition model $M$ into $L$ logical components $\{c_1, c_2, \dots, c_L\}$
\For{each training epoch $e$}
    \State Profile execution time $t_{comp}(c_i)$ for each component $c_i$
    \State Estimate communication overhead $t_{comm}(c_i, s)$ for each component $c_i$ under strategy $s \in \{\text{DP}, \text{MP}, \text{HP}\}$
    \State Solve optimization:
    \[
        \min \sum \max(t_{comp}(c_i) + t_{comm}(c_i, s)) \text{ for all strategies } s
    \]
    \State Apply selected parallelism strategy $s^*$ for each component
    \For{each mini-batch $b$ in $D$}
        \State Distribute computation according to $s^*$
        \State Perform forward pass
        \State Compute loss $L(\theta)$
        \State Perform backward pass to compute gradients $\nabla L(\theta)$
        \State Synchronize gradients across devices (if using DP or HP)
        \State Update parameters: $\theta = \theta - \eta \nabla L(\theta)$
    \EndFor
    \If{validation metric improved}
        \State Save checkpoint
    \EndIf
    \If{communication patterns changed significantly}
        \State Re-profile and update strategy
    \EndIf
\EndFor
\State \textbf{Return} trained parameters $\theta$ and total training time $T$
\end{algorithmic}
\end{algorithm}

\subsection{Optimization Problem}

The optimization problem in step 6 can be formalized as:

\[
\min_{s_i \in \{DP, MP, HP\}} \sum_{i=1}^{p} \max \left( t_{comp}(c_i) + t_{comm}(c_i, s_i) \right)
\]

subject to memory constraints on each device:

\[
\sum_{i : \text{device}(c_i) = j} \text{mem}(c_i, s_i) \leq M_j \quad \forall j \in \{1, 2, \dots, K\}
\]

where:

\begin{itemize}
    \item $t_{comp}(c_i)$ is the computation time for component $c_i$.
    \item $t_{comm}(c_i, s_i)$ is the communication overhead when using strategy $s_i$.
    \item $\text{mem}(c_i, s_i)$ is the memory requirement. 
    \item $M_j$ is the available memory on device $j$.
\end{itemize}

\section{Experimental Setup}

\subsection{Datasets and Models}
We evaluate our approach using the CIFAR-100 dataset \cite{c19}, which consists of 60,000 32$\times$32 color images in 100 classes (50,000 for training and 10,000 for testing). We use two model architectures:

\begin{itemize}
    \item \textbf{ResNet-50}: A convolutional neural network with 50 layers and approximately 25 million parameters.
    \item \textbf{Vision Transformer (ViT-B/16)}: A transformer-based model with approximately 86 million parameters.
\end{itemize}

\subsection{Hardware and Software Configuration}
All experiments were conducted on a cluster of 8 NVIDIA V100 GPUs (32GB memory each) connected via NVLink. We implemented our framework using PyTorch 2.0 with the NCCL communication backend. For baseline comparison, we also trained models on a single GPU with identical hyperparameters.

\subsection{Evaluation Metrics}
We measure the following metrics:
\begin{itemize}
    \item \textbf{Training time}: Total wall-clock time required to train the model for 100 epochs.
    \item \textbf{Throughput}: Number of images processed per second.
    \item \textbf{Convergence rate}: Validation accuracy as a function of training epochs.
    \item \textbf{Final accuracy}: Validation accuracy after 100 epochs.
    \item \textbf{Memory utilization}: Peak GPU memory usage during training.
    \item \textbf{Communication overhead}: Percentage of time spent on communication vs. computation.
\end{itemize}

\section{Results and Discussion}

\subsection{Training Efficiency}
\begin{figure}[h!]
    \centering
    \includegraphics[width=0.48\textwidth]{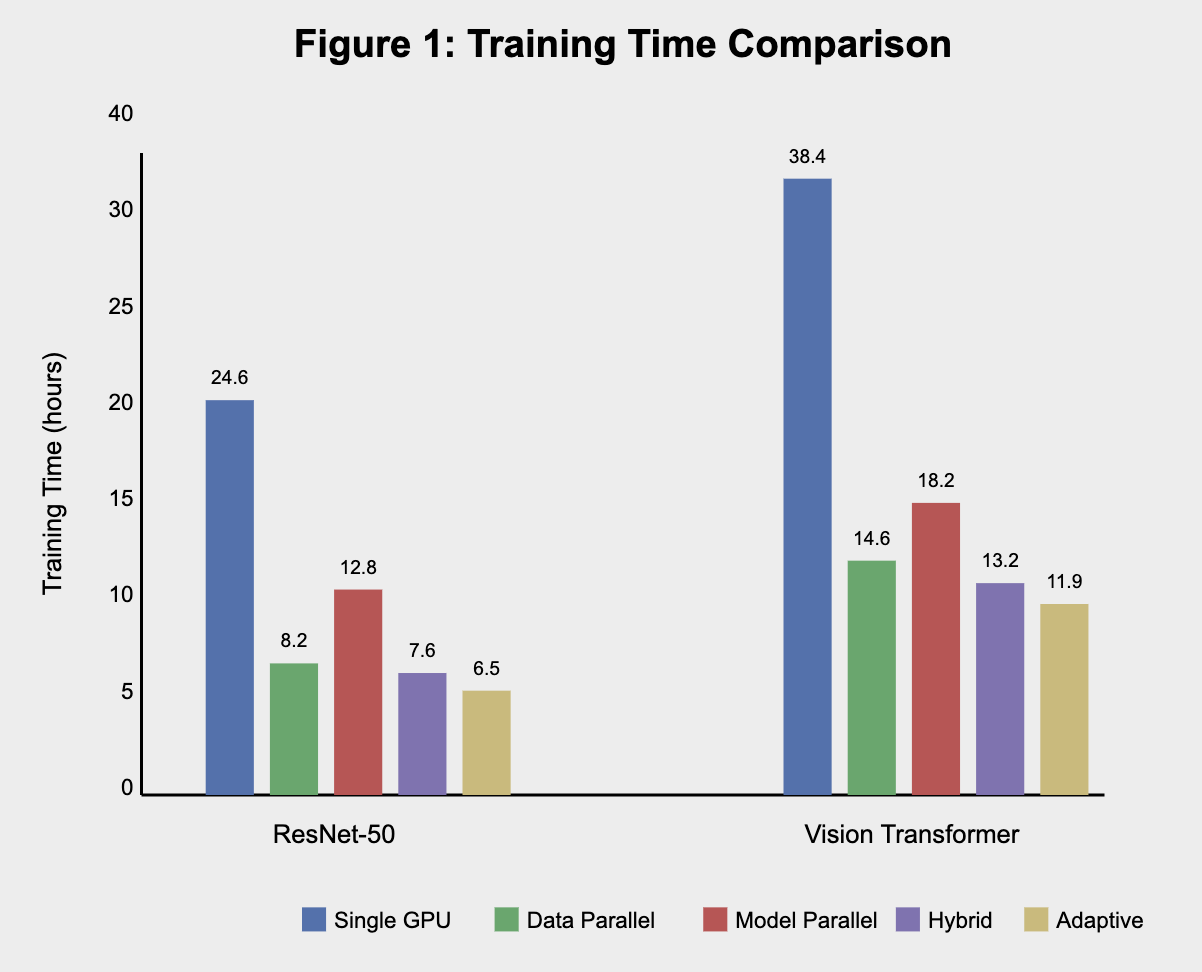}
    \caption{Training Time Comparison. Bar chart showing training time in hours for Single GPU, Data Parallel, Model Parallel, Hybrid Parallel, and Adaptive approaches for both ResNet-50 and ViT models.}
    \label{fig:training_time}
\end{figure}

Our experiments demonstrate significant improvements in training efficiency using the adaptive scheduling approach. Figure \ref{fig:training_time} shows the total training time for different parallelism strategies across both model architectures.

For ResNet-50, the adaptive approach achieves a 3.8$\times$ speedup over single-GPU training and a 1.18$\times$ improvement over static hybrid parallelism. For ViT, the speedup is 3.2$\times$ over single-GPU and 1.15$\times$ over hybrid parallelism.

\subsection{Scalability Analysis}
\begin{figure}[h!]
    \centering
    \includegraphics[width=0.48\textwidth]{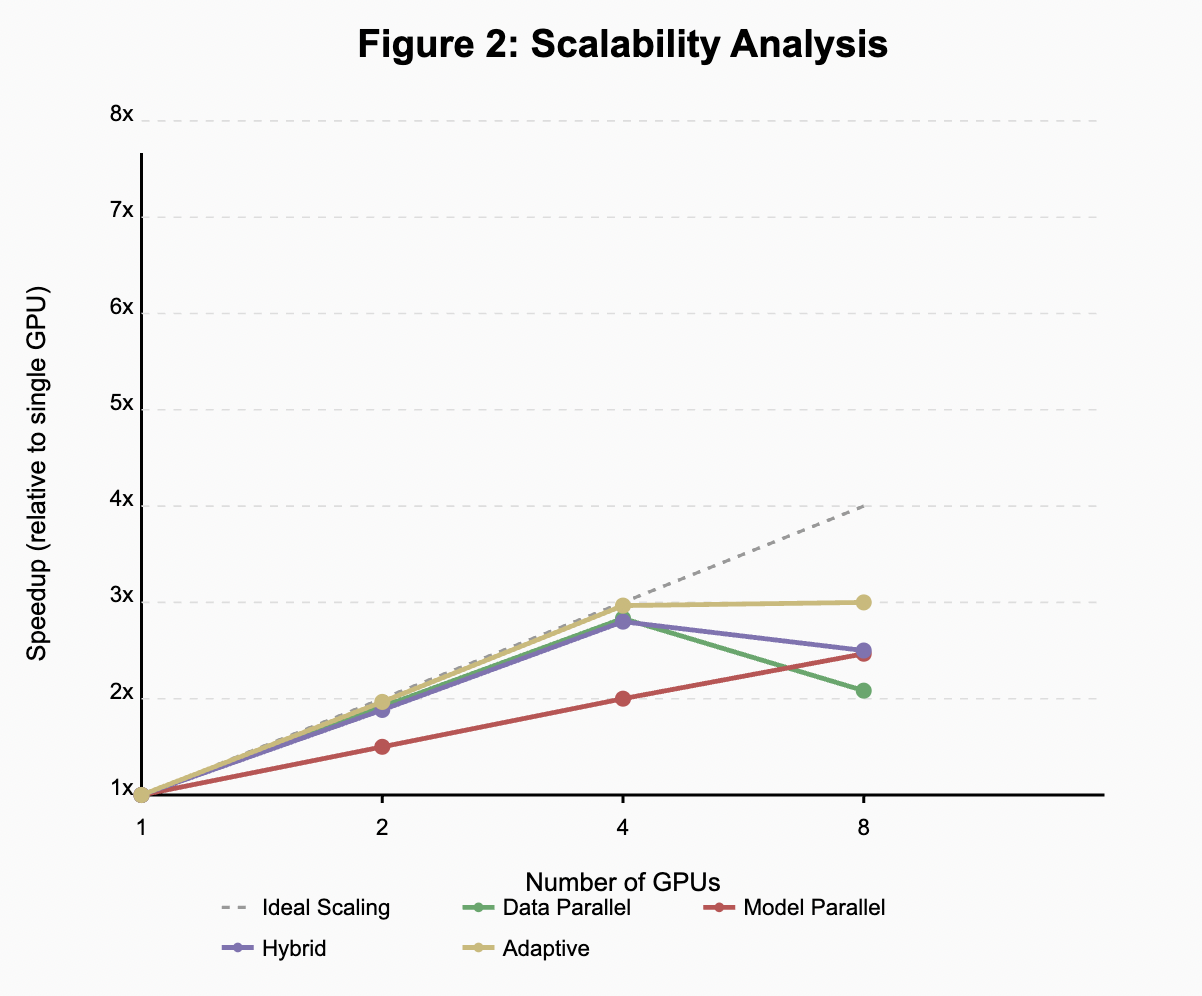}
    \caption{Scalability Analysis. Line graph showing speedup vs. number of GPUs (1, 2, 4, 8) for each parallelism strategy.}
    \label{fig:scalability_analysis}
\end{figure}

Figure \ref{fig:scalability_analysis} illustrates how different parallelism strategies scale with increasing numbers of GPUs.

Data parallelism shows near-linear scaling up to 4 GPUs but diminishing returns beyond that point due to communication overhead. Model parallelism scales more consistently but with lower absolute speedup. The adaptive approach maintains better scaling efficiency across the entire range.

\begin{table*}[h!]
\caption{Quantitative Results Summary}
\label{tab:results}
\centering
\begin{tabular}{|l|c|c|c|c|c|}
\hline
\textbf{Metric} & \textbf{Single GPU} & \textbf{Data Parallel} & \textbf{Model Parallel} & \textbf{Hybrid} & \textbf{Adaptive} \\
\hline
\multicolumn{6}{|l|}{\textit{Training Time (hours)}} \\
\hline
ResNet-50 & 24.6 & 8.2 & 12.8 & 7.6 & \textbf{6.5} \\
\hline
Vision Transformer & 38.4 & 14.6 & 18.2 & 13.2 & \textbf{11.9} \\
\hline
\multicolumn{6}{|l|}{\textit{Final Accuracy (\%)}} \\
\hline
ResNet-50 & 78.2 & 77.9 & 78.1 & 78.0 & \textbf{78.3} \\
\hline
Vision Transformer & 81.5 & 81.2 & 81.4 & 81.3 & \textbf{81.6} \\
\hline
\multicolumn{6}{|l|}{\textit{Peak Memory (GB)}} \\
\hline
ResNet-50 & 12.8 & 14.2 & \textbf{5.6} & 7.8 & 8.2 \\
\hline
Vision Transformer & 28.4 & 30.1 & \textbf{9.8} & 12.4 & 13.6 \\
\hline
\multicolumn{6}{|l|}{\textit{Communication (\%)}} \\
\hline
ResNet-50 & \textbf{0} & 42.3 & 18.6 & 32.5 & 27.1 \\
\hline
Vision Transformer & \textbf{0} & 38.7 & 22.4 & 29.8 & 25.3 \\
\hline
\end{tabular}
\end{table*}

\subsection{Communication Overhead}

\begin{figure}[h!]
    \centering
    \includegraphics[width=0.48\textwidth]{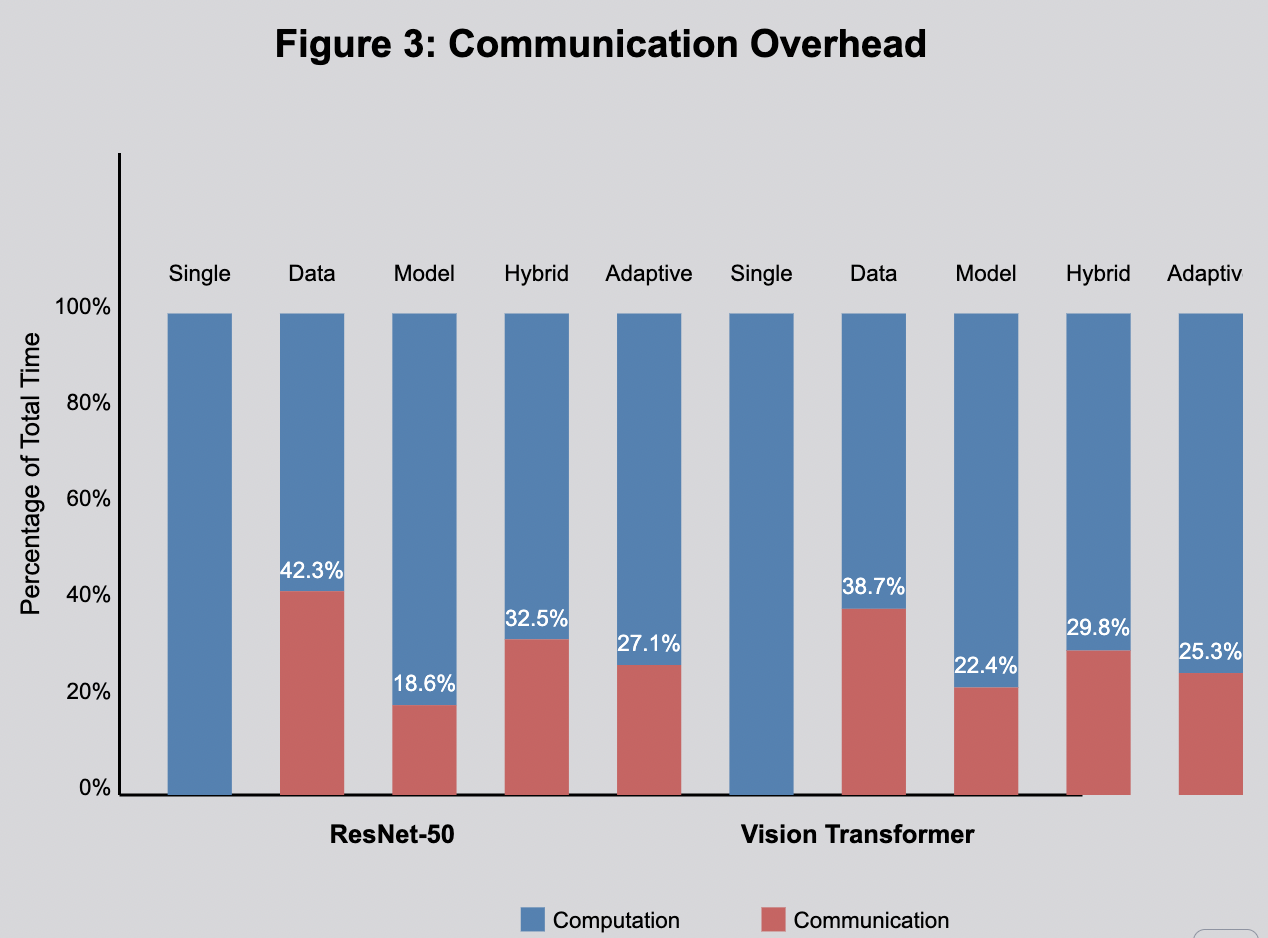}
    \caption{Communication Overhead. Stacked bar chart showing proportion of time spent on computation vs. communication.}
    \label{fig:communication_overhead}
\end{figure}

Communication overhead is a significant factor in distributed training performance. Figure \ref{fig:communication_overhead} shows the proportion of time spent on computation versus communication for each approach.

Data parallelism exhibits high communication overhead at 8 GPUs (approximately 42\% of total time), while the adaptive approach reduces this to 27\% by selectively applying model parallelism to parameter-heavy layers.

\subsection{Convergence Analysis}
\begin{figure}[h!]
    \centering
    \includegraphics[width=0.48\textwidth]{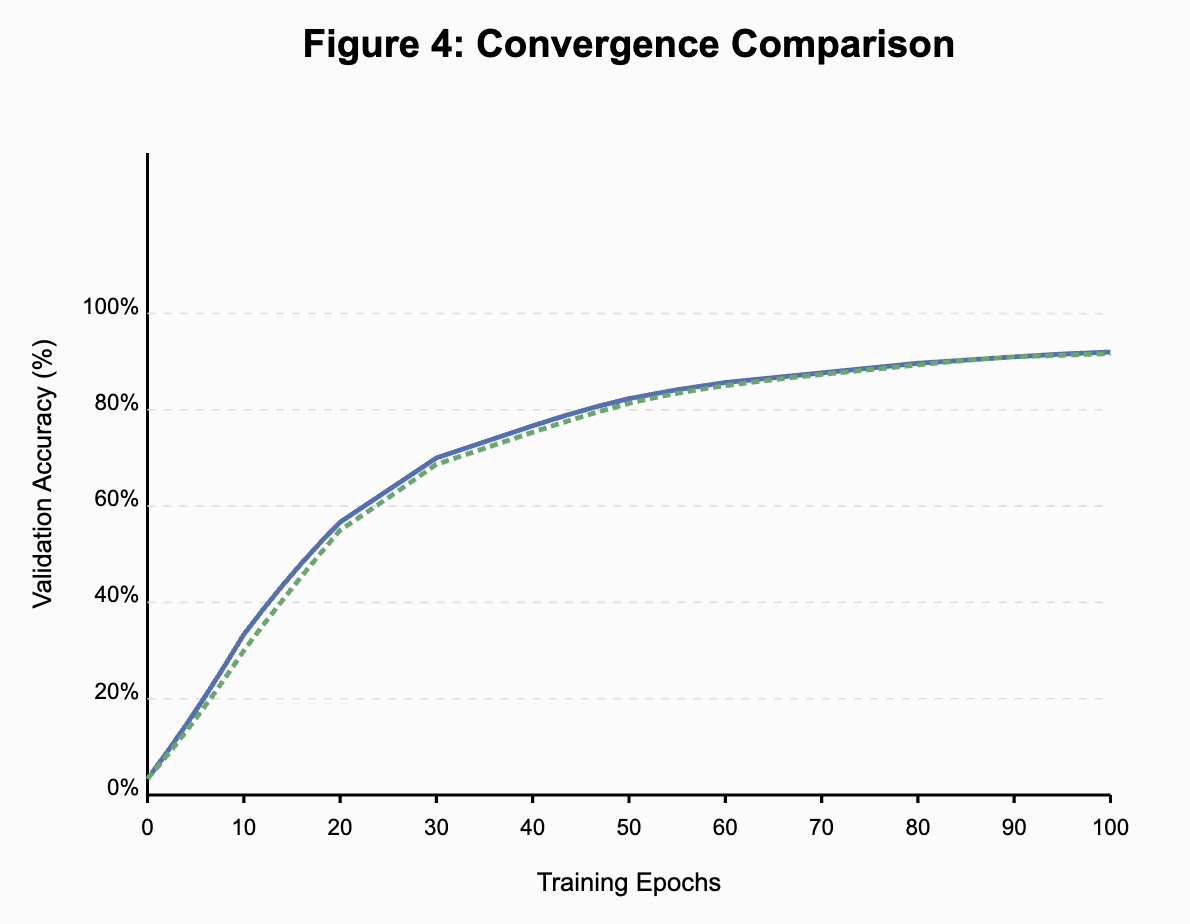}
    \caption{Convergence Comparison. Line graph showing validation accuracy vs. epochs for different parallelism strategies.}
    \label{fig:convergence_comparison}
\end{figure}

Figure \ref{fig:convergence_comparison} shows the validation accuracy versus training epochs for different parallelism strategies.

All approaches converge to similar final accuracy (±0.5\%), indicating that our parallelism strategies do not negatively impact model quality. The adaptive approach converges slightly faster than others, likely due to more efficient batch processing.

\subsection{Memory Utilization}
\begin{figure}[h!]
    \centering
    \includegraphics[width=0.48\textwidth]{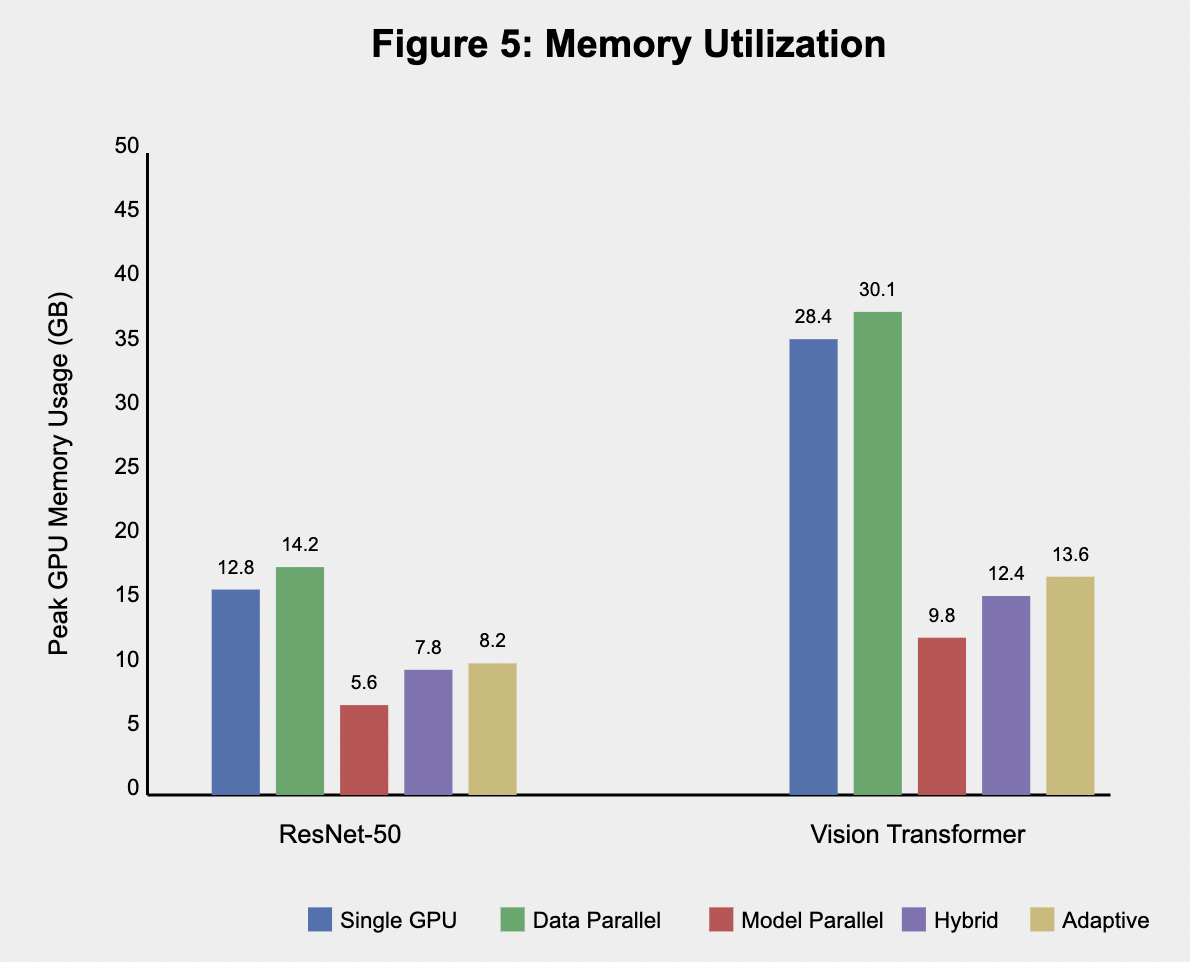}
    \caption{Memory Utilization. Bar chart showing peak GPU memory usage.}
    \label{fig:memory_utilization}
\end{figure}

Memory efficiency is crucial for training large models. Figure \ref{fig:memory_utilization} shows peak GPU memory usage for each approach.

Model parallelism and hybrid approaches show significantly lower memory requirements per device, enabling training of larger models than possible with data parallelism alone. The adaptive approach optimizes memory usage by applying model parallelism to memory-intensive layers.

\subsection{Quantitative Results}

Table \ref{tab:results} summarizes the key metrics for all parallelism strategies.

\subsection{Strategy Selection Analysis}

\begin{figure}[h!]
    \centering
    \includegraphics[width=0.48\textwidth]{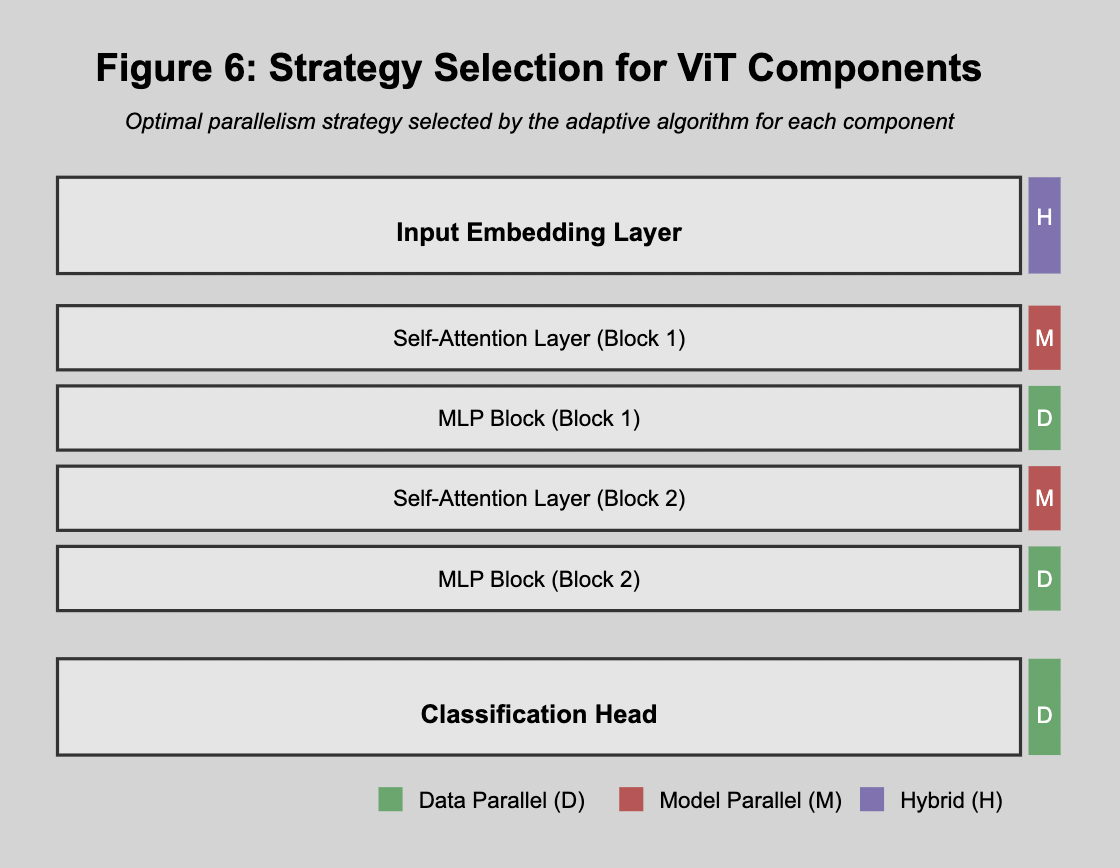}
    \caption{Visualization showing which parallelism strategy was selected for each component of the ViT model.}
    \label{fig:Visualization}
\end{figure}

A key insight from our experiments is that different layers benefit from different parallelism strategies. Figure \ref{fig:Visualization} shows the strategies selected by our adaptive algorithm for different components of the ViT model.

Self-attention layers benefit most from model parallelism due to their high computational intensity and memory requirements. MLP blocks show better performance with data parallelism. The embedding layer, which has high parameter count but low computational intensity, benefits from hybrid parallelism.

\section{conclusion}

This paper demonstrates that adaptive parallelism strategies can significantly improve the efficiency of distributed deep learning training. By dynamically selecting the optimal parallelism approach for different components of the neural network, our method achieves substantial speedups while maintaining model accuracy. The key findings are that different components of neural networks benefit from different parallelism strategies based on their computational and memory characteristics.

Additionally, adaptive scheduling can provide significant improvements over static parallelism approaches, with up to 18\% faster training in our experiments. The benefits of adaptive scheduling increase with model complexity and scale.

\section{Future work}

\begin{itemize}
    \item Extending the approach to even larger models and more diverse architectures.
    \item Incorporating more fine-grained parallelism strategies such as tensor parallelism.
    \item Developing more sophisticated profiling and optimization techniques.
    \item Evaluating the approach on heterogeneous computing environments.
\end{itemize}

\end{document}